\documentclass[aps,prl,showpacs,twocolumn,amsmath,amssymb]{revtex4}

\usepackage{revsymb}
\usepackage{graphicx}
\usepackage{dcolumn}
\usepackage{bm}

\begin{document}

\title{Pumping of nuclear spins by the optical solid effect in a quantum dot}

\author{E. A. Chekhovich$^{1,2}$, M. N. Makhonin$^{1}$, K. V. Kavokin$^{1,3}$,  A. B. Krysa$^4$, M. S. Skolnick$^1$, A. I. Tartakovskii$^1$}

\affiliation{$^{1}$Department of Physics and Astronomy, University
of Sheffield, Sheffield S3 7RH, UK\\
$^{2}$Institute of Solid State Physics, 142432, Chernogolovka, Russia\\
$^{3}$A. F. Ioffe Physico-Technical Institute, 194021, St. Petersburg, Russia\\
$^{4}$Department of Electronic and Electrical Engineering,
University of Sheffield, Sheffield S1 3JD, UK }

\date{\today}

\begin{abstract}

We demonstrate that efficient optical pumping of nuclear spins in
semiconductor quantum dots (QDs) can be achieved by resonant
pumping of optically ''forbidden'' transitions. This process
corresponds to one-to-one conversion of a photon absorbed
by the dot into a polarized nuclear spin, which also has potential
for initialization of hole spin in QDs.  Pumping via the ''forbidden'' transition is a
manifestation of the ''optical solid effect'', an optical analogue
of the effect previously observed in electron spin resonance
experiments in the solid state. We find that by employing this
effect, nuclear polarization of $65\%$ can be achieved, the
highest reported so far in optical orientation studies in QDs. The efficiency of the spin pumping exceeds that employing the allowed transition, which saturates due to the low probability of electron-nuclear spin flip-flop. 
\end{abstract}

\maketitle

Resonant optical pumping is a powerful method for direct control
of individual quantum states in a variety of physical systems 
\cite{Haffner,Santori} now also including semiconductor quantum dots \cite{Atature1,Mikkelsen,Gerardot,Press,Ramsay,Xu,Brunner,Xu1}.
In III-V semiconductor nano-structures the optically controlled
carrier spin dynamics are strongly influenced by the hyperfine
coupling of the electron spin with a bath of nuclei. The hyperfine
interaction (HI) is at the core of the resonant optical pumping
of the hole spin in charged dots
\cite{Gerardot,Brunner}. On the other hand, the HI shortens the
electron spin coherence and lifetime \cite{Khaetskii,KKM}
and therefore is unfavorable for electron spin manipulation in
nano-structures \cite{Xu1,Reilly,Petta}. Such undesirable effects
arising from fluctuations of the effective nuclear magnetic field 
can be suppressed by locking optical transitions in the dot to a resonant laser \cite{Xu1,Greilich}, or by pumping nuclear spins into a suitable narrow distribution of states \cite{Reilly,Burkard}. Another widely discussed approach is to pump very high nuclear spin polarizations \cite{Imamoglu}, a task so far eluding an experimental realization.

Here we present experiment and theory demonstrating that  excitation of a positively charged dot with light resonant with an optically forbidden transition leads to the most efficient nuclear polarization in QDs, exceeding that achievable by the well-studied optical spin pumping employing allowed transitions. This counter-intuitive observation sheds new light on the long standing problem of the realization of large nuclear polarization in semiconductors. Both non-resonant excitation and resonant pumping via the allowed transition, leading to the well-understood Overhauser effect, rely on high occupancy of a spin-polarized trion on the dot. They saturate with saturation of the optical transition at a relatively small spin pumping efficiency, limited by the low probability of the electron-nuclear spin flip-flop. In contrast, pumping via the spin-forbidden transition, enabled by electron spin state mixing due to the HI \cite{Korenev}, relies only on the presence of a hole. Its efficiency is only limited by the radiative recombination rate of the trion. This process, revealed in our work, arises from absorption of
a single photon accompanied by the simultaneous spin flip of an
electron and a single nucleus. This is a close analogue of the
''solid effect'' \cite{AbrahamBook,Bracker2}, a dynamic nuclear
polarization phenomenon observed in solids with paramagnetic
centers under microwave excitation.  By employing the ''optical
solid effect'' we find that nuclear polarization increases with
the laser intensity and saturates when the polarization degree reaches
65$\%$, the highest reported so far for QDs
\cite{Tartakovskii,Urbaszek,Skiba}. We show that this is
not a result of saturation of the optical transition itself, but
rather may be a limitation due to inherent properties of the
interacting electron-nuclear spin system, setting a
fundamental limit to the maximum polarization achievable by optical
pumping. The one-to-one photon-to-polarized-nucleus
conversion can also serve as a one-step initialization of the hole
spin.

The sample containing InP/GaInP QDs was grown by metal-organic
vapor phase epitaxy (see details in Ref. \cite{Skiba}). The
experiments were performed on individual dots at a temperature of
4.2~K, in external magnetic field $B_z$ up to 8~T perpendicular to
the sample surface. We study positively charged QDs emitting at
$\sim$1.84~eV. The sample was not intentionally doped. The dot
charging with holes, verified by magneto-spectroscopy and Hanle
effect measurements, occurred due to residual doping.

\begin{figure}
\includegraphics{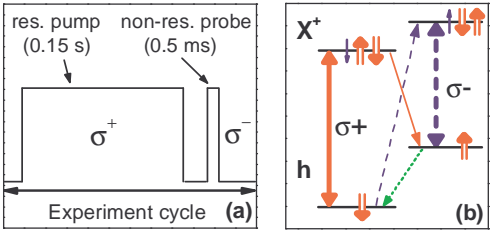}
\caption{\label{fig:Intro} (a) Pulse sequence used in the resonant
nuclear spin pump-probe experiment. (b) Energy level diagram of a
positively charged dot in magnetic field $B_z$. Electron and hole
spin up (down) states are shown by $\uparrow$($\downarrow$) and
$\Uparrow$($\Downarrow$) respectively. Long thick (thin) arrows
show ''allowed'' (''forbidden'') optical transitions. Dotted arrow shows hole spin relaxation.}
\end{figure}

The observation of nuclear spin pumping in this work has been realised using pump-probe techniques, where the effect of
the resonant pump excitation is determined by monitoring the whole
spectrum of the dot using a non-resonant probe, an important
contrasting factor to the selective spectral probing used in
absorption measurements
\cite{Atature1,Gerardot,Brunner,Xu1}. The pump-probe
experiment cycle is shown schematically in Fig.\ref{fig:Intro}(a).
A long circularly polarized pump pulse from a tunable single-mode
laser excites the dot. A short pulse from a second laser,
cross-circularly polarized relative to the pump and emitting below the
GaInP band-gap at $\sim$1.88~eV, is used to excite
photoluminescence (PL) from the positive trion $X^+$ and to probe
the states of the dot. The polarized PL excited by the probe pulse
is analyzed with a 1~m double spectrometer. This enables absolute
$X^+$ transition energies to be determined and Overhauser fields
for a given pump laser wavelength to be deduced. The duration of
the probe pulse was chosen to eliminate its effect on the nuclear spin
polarization on the dot: for most of the measurements we used
0.15~s and 0.5~ms pump and probe pulses, respectively. Multiple
cycles were repeated for each resonant laser frequency to improve
the signal to noise ratio, yielding a typical measurement time of
1~min, exceeding the time required for the nuclear polarization to
reach its steady-state value ($\tau_{build-up}\approx10$~s).

A diagram of the energy levels of the positively charged dot in
external magnetic field $B_z$ is shown in Fig.\ref{fig:Intro}(b).
The system ground state with hole spin up(down)
$\left|\Uparrow\right\rangle$($\left|\Downarrow\right\rangle$) and
the photo-excited $X^+$ trion state with electron spin up(down)
$\left|\Uparrow\Downarrow\uparrow\right\rangle$($\left|\Uparrow\Downarrow\downarrow\right\rangle$)
have energies $E_{\Uparrow(\Downarrow)}=\pm\mu_B g_h B_z/2$ and
$E_{\Uparrow\Downarrow\uparrow(\Uparrow\Downarrow\downarrow)}=E_{X^+}\pm\mu_Bg_e(B_z+B_{N})/2$
respectively, where $\mu_B$ is the Bohr magneton, $B_N$ is the
nuclear field, $g_{e(h)}\approx$+1.5(+3.1) is the electron(hole)
g-factor and $E_{X^+}$ is the PL energy at $B_z$=$B_N$=0. Allowed
$\sigma^{-(+)}$ polarized optical transitions from
$\left|\Uparrow\Downarrow\uparrow\right\rangle$($\left|\Uparrow\Downarrow\downarrow\right\rangle$)
to $\left|\Uparrow\right\rangle$($\left|\Downarrow\right\rangle$)
observed in PL are shown by thick arrows.

\begin{figure}
\includegraphics{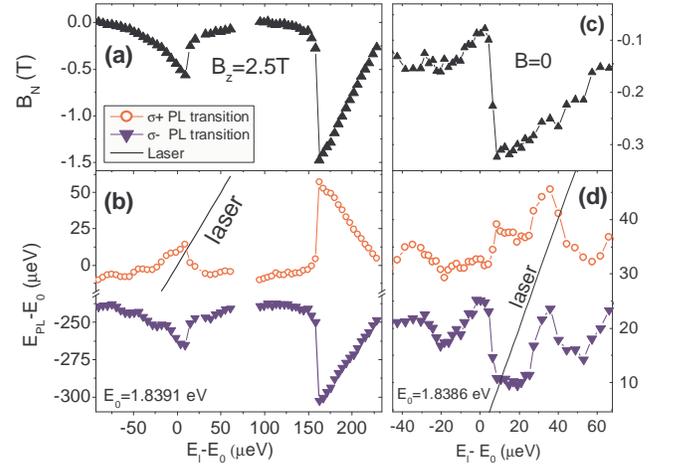}
\caption{\label{fig:ResScan} Overhauser field $B_N$ on the dot
(a,c) and PL transitions energies $E_{PL}$ (b,d) as a function of
the energy $E_l$ of the $\sigma^+$ polarized resonant laser with
excitation power $P_{res}$=15~$\mu$W at $B_z$=2.5~T (a,b) and
$B_z$=0 (c,d). The full lines on panels (b,d) show the laser
energy.}
\end{figure}

The effect of the resonant excitation with a $\sigma^+$ polarized pump laser is demonstrated in
Fig.\ref{fig:ResScan}. Here the magnitude of the nuclear field
$B_N$ deduced from the $X^+$ spectral splitting
$(E_{\Uparrow\Downarrow\downarrow}-E_{\Downarrow})-(E_{\Uparrow\Downarrow\uparrow}-E_{\Uparrow})$
is shown as a function of the pump laser energy $E_l$.

As shown in Fig.\ref{fig:ResScan}(a) for the case of high magnetic
field $B_z=2.5$~T and laser power $P_{res}=15~\mu$W, the
dependence of $B_N$ on laser frequency has the form of two
strongly asymmetric dips. When the laser is tuned from low energy
close to the resonance with the optically allowed transition
$\left|\Downarrow\right\rangle$$\leftrightarrow$$\left|\Uparrow\Downarrow\downarrow\right\rangle$,
$B_N$ decreases from 0~T to $-0.6$~T and then switches abruptly to
$B_N$$\sim-0.2$~T with a further slow increase to $\sim$0~T.
Surprisingly, another strong decrease of $B_N$ is observed when
the laser is tuned near the spin-forbidden transition
$\left|\Downarrow\right\rangle$$\leftrightarrow$$\left|\Uparrow\Downarrow\uparrow\right\rangle$
[thin dashed arrow in Fig.\ref{fig:Intro}(b)]. Here in marked
contrast, an abrupt change in $B_N$ from 0 to $-1.5$ T occurs at low
laser energy, with the maximum $|B_N|$ at this optical power 2.5
times larger than that excited via the spin-allowed transition.
When $E_l$ is tuned further, $B_N$ gradually increases back to 0.

The results in Fig.\ref{fig:ResScan}(a) are deduced from the
measurement of absolute $X^+$ PL transitions energies shown in
Fig.\ref{fig:ResScan}(b) as a function of laser energy. The
detuning between the $\sigma^+$ PL line and the laser at each
$E_l$ is given by the energy difference between the circles and the
black line in Fig.\ref{fig:ResScan}(b), which represents the energy
of the laser. In the case of the spin-allowed process the abrupt
change of nuclear polarization occurs just as the laser is tuned
slightly above the $\sigma^+$ PL line.

The same laser tuning experiment was repeated at $B$=0 [see Fig.
\ref{fig:ResScan}(c) and (d)]. At $B$=0 the shifts caused by
nuclear polarization are comparable to random spectral diffusion
of the dot PL energy, arising most likely from interaction with
random charges in the surrounding matrix. An asymmetric form of
the resonance similar to that in high magnetic fields can be seen
in Fig.\ref{fig:ResScan}(c). However, an abrupt change in $B_N$
occurs only when the $\sigma^+$ polarized laser comes into
resonance with the $\sigma^-$ polarized
$\left|\Uparrow\right\rangle$$\leftrightarrow$$\left|\Uparrow\Downarrow\uparrow\right\rangle$
PL transition coinciding with the forbidden
$\left|\Downarrow\right\rangle$$\leftrightarrow$$\left|\Uparrow\Downarrow\uparrow\right\rangle$
transition at $B$=0. Furthermore, the asymmetry
of the resonance curve is similar to that observed for pumping via
the spin-forbidden transition at $B_z=2.5$~T in
Fig.\ref{fig:ResScan}(a): the abrupt change in $B_N$ is observed
towards lower laser energy.

In order to explain the observed effects we consider the resonant nuclear spin pumping mechanism illustrated in
Fig.\ref{fig:Intro}(b). As the QD contains of the order of 10$^4$
magnetic nuclei, their polarization requires many cycles of
optical excitation/recombination. The optically allowed circularly
polarized transitions shown in Fig.\ref{fig:Intro}(b) with thick
arrows conserve spin and cannot be responsible for nuclear spin
pumping. For the $\sigma^+$ polarized excitation used in
Fig.\ref{fig:ResScan}, there are two possible cyclic processes
yielding nuclear polarization. Both start from the spin-down state
of the resident hole $\left|\Downarrow\right\rangle$ and involve
optically forbidden transitions:

(1) The cycle associated with the Overhauser effect
(OE) consists of the following steps (i) excitation of the
allowed transition to the spin-down
$\left|\Uparrow\Downarrow\downarrow\right\rangle$ state [thick
solid arrow in Fig. \ref{fig:Intro}(b)]; (ii) recombination to the
spin-up hole state $\left|\Uparrow\right\rangle$ (thin solid
arrow), assisted by the hyperfine interaction (otherwise this
transition is forbidden); (iii) hole spin-flip
back into $\left|\Downarrow\right\rangle$ state (dotted arrow).

(2) The cycle associated with the optical solid effect (SE) also
involves three steps: (i) hyperfine-interaction-assisted
excitation of the spin-up
$\left|\Uparrow\Downarrow\uparrow\right\rangle$ state via the
spin-forbidden transition (thin dashed arrow); (ii) X$^+$
recombination via the allowed transition into the
$\left|\Uparrow\right\rangle$ state (thick dashed arrow); (iii)
and, finally, the hole spin-flip completing the cycle.

Each of the above excitation cycles results in change of
the $Z$-projection of the total nuclear spin in the QD by~$-1$. We
calculate the total rate of spin pumping into the nuclear system
$W_{pump}$ as the sum of the rates of these cycles
by solving optical Bloch equations for the two photo-excitation
paths.

\begin{figure}
\includegraphics{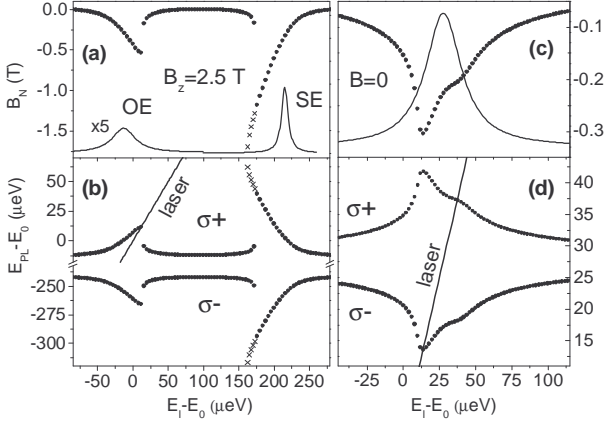}\caption{
\label{fig:TheorResScan} (a) and (c) show calculated nuclear field
$B_N$ on the dot ([crosses]circles show [meta]stable solutions)
and nuclear spin pumping rate $|W_{OE}+W_{SE}|$ at $B_N$=0
(lines). (b) and (d) show calculated PL transitions energies
$E_{PL}$ as a function of the energy  $E_l$ of the $\sigma^+$
polarized laser at $B_z$=2.5 (a,b) and $B$=0~T (c,d). Lines on
panels (b,d) show the laser energy.}
\end{figure}

For the SE cycle, the lifetime $\tau$  of the excited state is
determined by the allowed recombination transition. In contrast,
for the OE cycle the recombination rate into the spin-up hole
state is reduced by the factor $\alpha^2=\mu^2_B g^2_e
\langle\delta B^2_N\rangle/E^2_{eZ}$ characterizing mixing of the
two X$^+$ states by the hyperfine interaction. Here $\langle\delta
B^2_N\rangle$ is the mean squared fluctuation of the Overhauser
nuclear field, while
$E_{eZ}=E_{\Uparrow\Downarrow\uparrow}-E_{\Uparrow\Downarrow\downarrow}$
is the electron Zeeman splitting characterizing the energy
difference between X$^+$ trion states. The hyperfine admixture
factor $\alpha$ also governs the optical matrix element for the
excitation transition in the SE cycle, which is given by
$\alpha\Omega_R$  rather then $\Omega_R$  as in the case of the OE
cycle (here $\Omega_R$  is the laser Rabi frequency). Note that in
our experiments the splitting of the X$^+$ states always exceeds by far
the magnitude of the electron splitting due to the nuclear
spin fluctuation $\mu_B g_e \langle\delta
B^2_N\rangle^{1/2}\approx1~\mu$eV, and therefore $\alpha$ remains
small. As a result, we obtain the following expressions for the
rates of OE and SE cycles:
\begin{eqnarray}
\label{eq:wpump}
W_{OE}=-\frac{\alpha^2\tau^{-1}\cdot(\tau\Omega_R)^2/2}{1+(\tau\Omega_R)^2+(E_{\Uparrow\Downarrow\downarrow}-E_{\Downarrow}-E_{l})^2\tau^2\hbar^{-2}}\nonumber\\
W_{SE}=-\frac{\tau^{-1}\cdot(\alpha\tau\Omega_R)^2/2}{1+(\alpha\tau\Omega_R)^2+(E_{\Uparrow\Downarrow\uparrow}-E_{\Downarrow}-E_{l})^2\tau^2\hbar^{-2}}.
\end{eqnarray}
At low pumping ($\tau\Omega_R\ll$1) both
processes result in identical resonance lineshape, shifted in energy
by the splitting of the $X^+$ states. However, with
increasing pump intensity ($\tau\Omega_R\approx$1), $W_{OE}$
saturates and broadens, while $W_{SE}$ continues to grow linearly with
pump intensity up to $\tau\Omega_R\approx\alpha^{-1}\gg$1.
Thus the overall efficiency of the SE process can exceed that of the OE by
a factor of $1/\alpha^2$.

The stationary value of the nuclear field is found from the
condition that the total pumping rate $W_{pump}$ via SE and OE
processes must equal the loss rate of nuclear polarization:
\begin{eqnarray}
\label{eq:rateeq} W_{OE}(B_N,E_l)+W_{SE}(B_N,E_l)=\gamma
N\frac{B_N}{B_{N,max}}
\end{eqnarray}
where $\gamma$ is the rate of nuclear spin loss via spin
relaxation and diffusion, $N$ is the number of nuclei in the QD,
$B_{N, max}$ is the maximum $|B_N|$ corresponding to 100$\%$
polarized nuclear spins ($\mu g_e B_{N, max}$=230~$\mu
eV$ in InP \cite{Gotschy}).

Numerical solutions of the nonlinear equation Eq. \ref{eq:rateeq}
for values of parameters relevant to our experiment are shown in
Fig.~\ref{fig:TheorResScan}. The Overhauser field $B_N$ (dots) and
nuclear spin pumping rate $|W_{pump}|$ at $B_N$=0 are shown in
panels (a) and (c) for $B_z=2.5$~T and $B$=0 respectively.
Calculated PL transition energies
$E_{\Uparrow\Downarrow\downarrow}-E_{\Downarrow}$ and
$E_{\Uparrow\Downarrow\uparrow}-E_{\Uparrow}$ are shown in panels
(b) and (d) with symbols, and the pump laser energy is shown by
the continuous lines.

The best description of the experiment is achieved with realistic
parameters: nuclear polarization fluctuation $\mu_B g_e
\langle\delta B^2_N\rangle^{1/2}\approx 1.53(1.05)~\mu$eV,
electron spin level broadening $\hbar/\tau\approx$6.8(5.5)~$\mu$eV
and excitation power is $\Omega_R\tau\approx$2.0(3.5) at
$B_z$=0(2.5)~T. As expected, the best fits reflect the strong
suppression of the nuclear spin depolarization rate in high
magnetic fields: $1/(\gamma N)\approx$2~$ns$ at $B$=0 and
increases up to $\approx1.4$~$\mu s$ at $B_z$=2.5~T.

At high magnetic field the calculated behavior is in
excellent agreement with the experimental data in
Fig.\ref{fig:ResScan}a,c: under $\sigma^+$ polarized
excitation switching to low $|B_N|$ occurs towards high $E_l$, when
$E_{l}=E_{\Uparrow\Downarrow\downarrow}-E_{\Downarrow}$,
corresponding to the $\sigma^+$ PL (allowed) transition. This
process is described by $W_{OE}$ in Eq. \ref{eq:rateeq}. As in experiment another
abrupt switching to a markedly higher nuclear polarization is
observed when the laser is tuned to the forbidden
$\left|\Downarrow\right\rangle$$\rightarrow$$\left|\Uparrow\Downarrow\uparrow\right\rangle$
transition, the process described by $W_{SE}$.

The origin of the asymmetric shapes of the resonances in
Figs.\ref{fig:ResScan} and \ref{fig:TheorResScan} lies in the
non-linear dependence of the trion transition energies on the
nuclear field $B_N$. As a consequence both the detuning from the
laser and the nuclear spin pumping rate $W_{pump}$ are also
dependent on $B_N$. The highest nuclear spin pumping efficiency in
both OE and SE processes is reached when the laser is in exact
resonance with the
$\left|\Downarrow\right\rangle$$\rightarrow$$\left|\Uparrow\Downarrow\downarrow\right\rangle$
and
$\left|\Downarrow\right\rangle$$\rightarrow$$\left|\Uparrow\Downarrow\uparrow\right\rangle$
transitions, respectively. If in this condition the laser is tuned
away from resonance $|B_N|$ can only decrease. This leads to the
shift of the $\left|\Uparrow\Downarrow\downarrow\right\rangle$
state involved in the OE process to lower energy. If the laser is
tuned above resonance this shift leads to even larger detuning and
lower nuclear spin pumping efficiency. Such feedback
results in a quick collapse of $|B_N|$. If the laser is tuned to
lower energy, the shift of the
$\left|\Uparrow\Downarrow\downarrow\right\rangle$ level partly
restores the resonant condition making the reduction in the nuclear
spin pumping efficiency more gradual. In contrast to the OE case,
the $\left|\Uparrow\Downarrow\uparrow\right\rangle$ trion state
involved in the SE shifts to lower energy with the build-up of negative $B_N$. Thus the asymmetric shape of the SE resonance
exhibits the sharp drop in $|B_N|$ in the direction of low $E_l$
\cite{metastable}.

Calculations for the case of $B$=0
(Fig.\ref{fig:TheorResScan}~c,d) also show an asymmetric resonance
with the abrupt change in nuclear polarization on the low energy
side, in agreement with the results in Fig.\ref{fig:ResScan}. As
in experiment, the abrupt increase of $|B_N|$ under $\sigma^+$
excitation is observed when the laser is in exact resonance with
the $\sigma^-$ PL line (transition to
$\left|\Uparrow\Downarrow\uparrow\right\rangle$ state),
demonstrating efficient nuclear spin pumping through the forbidden
optical transition.

According to Eq.\ref{eq:wpump} $W_{OE}$ reaches a maximum value of
$W^{max}_{OE}=\alpha^2\tau^{-1}/2$ when the allowed transition
saturates ($\tau\Omega_R\gg$1). By contrast, the maximum
$W^{max}_{SE}=\tau^{-1}/2$ is limited only by the radiative
lifetime of the trion, $\tau$, and exceeds $W^{max}_{OE}$ in a
wide range of magnetic fields by $\alpha^{-2}>3000$ \cite{alpha}.
According to Eq.\ref{eq:wpump} $W_{SE}$ also saturates at much
higher laser power ($\alpha\tau\Omega_R\gg$1) than $W_{OE}$.
By performing power-dependent measurements, we find that the maximum 
nuclear polarization pumped via the allowed transition increases with power
and saturates at 45$\%$. For pumping via the ''forbidden'' transition, we obtain a maximum nuclear polarization of $65\%$, 1.5 and 1.2 times higher than the saturation levels observed for the OE cycle and non-resonant excitation, respectively. Surprisingly, the saturation of the pumping via the ''forbidden'' transition occurs
at a similar power to the OE case.
We thus conclude that the saturation of $|B_N|$ is not due to saturation of the forbidden transition itself. These results of the power-dependence (to be reported in detail elsewhere) suggest that the complete physical picture of the observed phenomena should also include additional mechanisms such as, for example, decreasing availability of unpolarized nuclei or formation of dark nuclear states \cite{Imamoglu,Burkard} in the dot. Such processes, yet to be fully understood, may set a fundamental limit to the efficiency of the optical orientation of nuclear spins.

In conclusion, we have employed resonant pump-probe techniques to
demonstrate the ''optical solid effect'' based on optically forbidden transitions in 
a positively charged quantum dot. This phenomenon results in efficient nuclear polarization over a wide range of finite and zero magnetic fields. Using model calculations based on optical Bloch equations we show that interaction of the
laser with both allowed and forbidden QD transitions is significant, and find that in realistic conditions the forbidden process is the most efficient mechanism of dynamic nuclear polarization.

Since submission of our manuscript, related work on resonant pumping of
negatively charged dots has been published \cite{Latta}.

We are grateful to V. I. Fal'ko, A. J. Ramsay, V. L. Korenev and V. D. Kulakovskii for
fruitful discussions. This work has been supported by the EPSRC
grants EP/G601642/1, EP/C54563X/1, the EPSRC IRC for QIP, and the
Royal Society.


\begin{thebibliography}{10}
\bibitem{metastable} Metastable nuclear spin states are observed in experiment and will be reported elsewhere. Relevant theory results are shown in Fig.\ref{fig:TheorResScan}(a).
\bibitem{alpha} The forbidden transition was not observed in PL measurements, which taking into account the signal-to-noise ratio in PL provides the upper estimate for $\alpha^2<1/3000$.
\end{thebibliography}

\end{document}